\documentclass[aps,superscriptaddress,showkeys]{revtex4}

\usepackage[dvips]{graphicx}

\begin{document}

\title{Phase Transitions In Compact Stars%
\thanks{student research project: Ann\'ee 2002-2003, DSM 2A, option
  physique}}

\author{Jean Macher}

\email{jmacher@ens-lyon.fr}

\affiliation{
Ecole Normale Sup\'erieure de Lyon,
46, All\'ee d'Italie,\\
F-69364 LYON Cedex 07, 
France}

\affiliation{
Institut f\"ur Theoretische Physik, 
J. W. Goethe Universit\"at,
Robert-Mayer-Stra\ss{}e 10,
D-60325 Frankfurt am Main,
Germany}

\author{J\"urgen Schaffner-Bielich}

\email{schaffner@astro.uni-frankfurt.de}

\affiliation{
Institut f\"ur Theoretische Physik, 
J. W. Goethe Universit\"at,
Robert-Mayer-Stra\ss{}e 10,
D-60325 Frankfurt am Main,
Germany}

\begin{abstract}
  We report on a three--month research project for undergraduate
  students about the mass-radius relation of compact stars. The equation
  of state used is constrained at low densities by well-established
  equations of state of the nuclear phase (the solid crust) and then
  extended to higher densities with a phenomenological, parametric
  approach. A first order phase transition from hadronic matter to a
  phase of higher density, assumed to be quark matter is studied in
  addition. The mass-radius relation is obtained by solving numerically
  the Tolman-Oppenheimer-Volkoff equation. We derive some conditions for
  the existence of a third family of compact stars on the form of the
  equation of state and its different global properties.
\end{abstract}

\keywords{neutron star, neutron drip, hadron, quark, first order phase
  transition, equation of state, mass-radius relation}

\maketitle

\section{Introduction}

The study of compact stars begins with the discovery of white dwarfs
(WD), and the successful description of their properties by the
Fermi-Dirac statistics, assuming that they are held up against
gravitational collapse by the degeneracy pressure of the electrons, an
idea first proposed by Fowler in 1926 \cite{Fowler26}. A maximum mass
for white dwarfs was found to exist in 1930 by the seminal work of
Chandrasekhar \cite{chandra31} due to relativistic effects. In 1932
Chadwick discovered the neutron. Immediately, the ideas formulated by
Fowler with electrons were generalised to neutrons, and the existence of
a new class of compact star, with a large core of degenerate neutrons,
was predicted: neutron stars (thereafter NS) \cite{Landau32}. The first
NS model calculations are achieved by Oppenheimer and Volkoff in 1939,
describing the matter in the star as an ideal degenerate neutron gas.
Their calculations showed the existence of a maximum mass, like in the
case of white dwarfs, above which the star is not stable and collapses
into a black hole. They found a maximum stable mass of $0.75 M_{\odot}$
(see \cite{OV39}). In the case of NS's, on the contrary to WD's, the
theory is far ahead of the observations. Only nearly 40 years later, in
1967, the first neutron star is observed --- in fact a strange object
pulsating in the radio range (a pulsar), but quickly identified as a
fast rotating NS.

In 1974, a pulsar (PSR 1913+16) is observed for the first time in a
binary system by Hulse and Taylor. It allowed a precise measurement of
its mass, which was found to be $1.44 M_{\odot}$. Hence, this mass
measurement rules out the simple picture of an ideal gas of neutrons for
the matter inside the star, and shows that the interactions between
the nucleons must be taken into account.

Even before the discovery of pulsars as rotating neutron stars, and
shortly after the introduction of the quark model, theoreticians
speculate about the possible existence of quark matter inside neutron
stars \cite{Ivan65}. Gerlach demonstrates in his PhD thesis with Wheeler
in 1968, that a third family of compact stars can possibly exist in
nature, besides white dwarfs and neutron stars \cite{Gerlach68}. He
derives general conditions on the equation of state for such a new form
of compact stars to exist, in particular that a strong softening of the
equation of state, like in a phase transition, has to occur in the
interior of neutron stars.  Heintzmann, Hillebrandt and coworkers
follow up the idea of Gerlach with first calculations using
Pandharipande's EOS with a phase transition to hyperons (see
\cite{Heintzmann74} and references therein). A more detailed
investigation for the existence of the third family is performed by
K\"ampfer motivated by the possible existence of a pion condensate or
quark matter inside compact stars. He parametrised the EOS in polytropic
form and derived some limits on the phase transition for the existence
of the third family \cite{kaempfer81,kaempfer83}.

Some astrophysicists even argue that the very ground state of matter is
in fact strange quark matter (composed of the quarks u, d and s), so
that any neutron star should end in an object containing only strange
quark matter. Such objects are studied since the mid 80's (see
\cite{Haensel86,Alcock86}) and are now referred to as \emph{strange stars}.

The investigation of the third family of compact stars is recently
rejuvenated by several works which calculate the equation of state in
relativistic field theoretical models and the phase transition to exotic
matter using Gibbs criteria (see
\cite{Glen_book,GK2000,Schertler00,FPS01,Banik01,Scha02,Mishustin03,Banik03}).
The approaches used are strikingly different in all these calculations,
in particular the exotic phase studied in the core of the neutron star
are quark matter, hyperon matter and Kaon condensed matter.
Nevertheless, a third family is always found if the phase transition is
sufficiently strong.

The lack of knowledge in the description of ultra-dense cold matter, in
particular above normal nuclear density, is a strong obstacle in the
study of the detailed properties of neutron stars and the possible
existence of the third family in particular. On the other hand, a
parametric approach to the problem not only simplifies the calculations
but gives a more general handle on the underlying parameters of the
equation of state.  The work by K\"ampfer mentioned above is a strong
motivation for a student project to readdress the issue of the third
family of compact stars in this approach taking into account the recent
advances in the field.

In the summer of 2003, Macher starts this student research project as a
part of his second year of the Magist\`ere des sciences de la mati\`ere
of the ENS-Lyon. Schaffner-Bielich at the Goethe University of
Frankfurt has the pleasure to propose and to guide him through the
research topic. The chosen scientific research area is at the forefront
of present active research on neutron stars, while being tractable at
the undergraduate level. In particular, basic knowledge about
thermodynamics, quantum statistics and relativity theory is sufficient
to tackle the scientific problem. We refer the reader to the excellent
textbooks about compact star physics \cite{Shapiro_book, Glen_book,
  Weber_book}. After finishing this work, the article by Silbar and
Reddy about neutron stars for undergraduates appeared \cite{Silbar04}
which with the above mentioned textbooks can be used as baselines for
student projects on compact stars (see \cite{Jackson04}).

The paper is outlined as follows: the general properties of neutron star
and astrophysical constraints as well as the rough structure of neutron
stars are discussed in the first section \ref{sec:properties}. The
parametric approach to the equation of state is delineated in section
\ref{sec:eos} and results are given in section \ref{sec:results}.

\section{General properties of neutron stars}
\label{sec:properties}

\subsection{Mechanical structure of a compact star}

\subsubsection{Arguments for using the general relativity theory}

We will be concerned mostly, throughout this work, with neutron stars
and more compact stars. The typical mass is of order $\sim 1 M_{\odot}$,
and the radius of order $10$ km, that is $10^{-5} R_{\odot}$, and thus
the gravitational field at the surface growing proportional to
$1/R^{2}$, the magnitude of the gravitational field at the surface of a
neutron star will be $10^{10}$ times that of the Sun. It's clear that in
those conditions, the curvature of space-time cannot be ignored, and
that neutron stars and \emph{a fortiori} hypothetical more compact stars
must be described within the framework of general relativity.

\subsubsection{Equations for the structure}

We consider the following system: a perfect fluid, spherically symmetric,
at rest. We use spherical coordinates $(r,\theta,\varphi)$. We denote
the pressure at a radius $r$ by $P(r)$, and the \emph{energy density},
containing all possible contributions except for the gravitational
energy, by $\rho(r)$ (in fact for a perfect fluid, $\rho$ is simply the
classical mass density multiplied by $c^{2}$). With these assumptions,
Einstein's equations yield the following first order differential
equations, the first one being known as the Tolman-Oppenheimer-Volkoff
(or TOV) equation \cite{Tolman39,OV39} (we take $c = 1$):
\begin{eqnarray}
\label{tov} \frac{dP}{dr}\quad &=& \quad -\frac{G\rho
m}{r^{2}}\frac{\Big(1+\frac{P}{\rho}\Big)\Big(1+\frac{4 \pi P
r^{3}}{m}\Big)}{\Big(1-\frac{2Gm}{r}\Big)} \\
\frac{dm(r)}{dr}\quad &=& \quad 4\pi r^{2}\rho(r)
\end{eqnarray}
The second equation defines simply the quantity $m(r)$, which is
naturally introduced by solving Einstein's equations, and represents
simply the total energy contained in the sphere of radius $r$. Thus at
$r=0$, $m$ must be zero and at $r=R$, $m$ is the total energy or mass $M$ of
the star. The derivation of the TOV-equation can be found in standard
textbooks (see e.g.\ \cite{Adler65,Weinberg72,Misner73,Glen_book,Weber_book}).
We note, that an anisotropy in the pressure may appear at nuclear
densities which leads to an extra term in the TOV equations
\cite{Herrera97}. This anisotropy can have possible impacts on the
maximum mass and the stability of neutron stars, which we will ignore in
the following.

The factor $-G\rho m/r^{2}$ on the right-hand side of equation \ref{tov}
is nothing else than the term from the classical equation for
hydrostatic equilibrium for a perfect fluid. Thus, the TOV equation
appears to be the general relativistic version of hydrostatic
equilibrium, where the Newtonian term is multiplied by relativistic
correction factors arising from the curvature of space-time due to the
presence of a finite energy density $\rho$. The unknowns in these two
equations are $\rho$, $P$, and $m$. We must thus have a third equation
to close the system. This third equation is the equation of state
$P=P(\rho)$, which contains the micro-physics of the matter in the star.

\subsubsection{Solving the equations}

The right-hand side of the TOV equation is always negative. Hence, the
pressure in the star decreases monotonically from its center to its
surface. The above equations are thus solved numerically from $r = 0$,
until the pressure vanishes. The radius at which the pressure vanishes
is the radius of the star $R$. The initial condition for $m(r)$ is
fixed: $m(0) = 0$, because the central point contains no energy. We
choose the value of the central pressure or central energy density. We
thus get the profiles $P(r)$, $\rho(r)$ (through the equation of state),
$m(r)$, and the value of the radius of the star, $R$, and the total
energy $M = m(R)$. By solving the equations for several values of the
central energy density, we obtain as many values of $M$ and $R$,
constructing a mass-radius relation.

\subsubsection{Importance of knowing the maximum mass}

The precise knowledge of the maximum mass (and radius) of a family of
stars allows the identification of objects observed. At the
moment, all we know for sure about an object with $10 M_{\odot} \geq M
\geq 1 M_{\odot}$ and $R < 50 km$ is that it is either a neutron star,
or a quark star, or a black hole. If we know the maximum mass $M_{max}$
of the neutron star family, then an object with $M > M_{max}$ is
necessarily a black hole or a quark star. To be complete, we also need
the mass-radius relation for the hypothetical third family, and the
confirmation of its existence, so that we can firmly distinguish black
holes from compact stars, and compact stars between them. To distinguish
objects with $M < M_{max}$, the precise knowledge of the entire
mass-radius relation and of the radii of these objects is needed.

Then, knowing what kind of object we are observing, we can use the
observed properties to constrain the models for the very badly known
dense and cold matter, and to rule some out. For a simple, enlightening
review about the maximum mass of neutron stars, see \cite{haensel03}.

\subsection{The states of matter in a neutron star: what we do know, and
  what we don't} 

The energy density in a neutron star ranges from $10^{14}$ to $10^{15}
g\cdot cm^{-3}$ at the center (the value in $g\cdot cm^{-3}$ is simply
the energy density divided by $c^{2}$), to zero at the surface, so that
almost every thinkable state of matter can be found.  Let us make a
quick list of the different phases in a neutron star and the phenomena
involved.  More details can be found in Appendix A. The thickness
of each phase given is valid for a typical neutron star, with a mass $\sim 1
M_{\odot}$.

\begin{figure}
\begin{center}
\includegraphics[scale = 0.75]{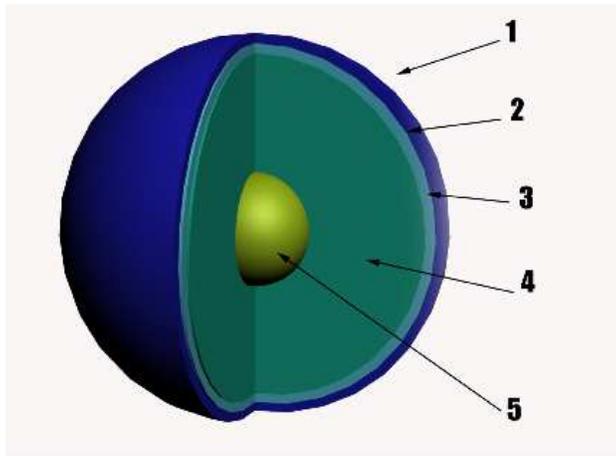}
\caption{The different layers in a neutron star (details in the 
    text). The star is represented at scale and has a radius of
    10 km, the outer crust and the inner crust are both 500 m thick, the
    inner core has a 3 km radius.}
\label{nslayer}
\end{center}
\end{figure}

\begin{enumerate}
\item Outer crust: $10^{4}\leq \rho \leq 10^{11} g\cdot cm^{-3}$, a few
  hundred meters.  \newline Lattice of neutron-rich nuclei, in a free
  gas of (relativistic) electrons.
\item Neutron drip line. At this point ($\rho \simeq 4.3\cdot 10^{11}
  g\cdot cm^{-3}$), the neutrons in the nuclei are so weakly bound that
  they ``drip'' out of the nuclei, progressively becoming free.
\item Inner crust: $10^{11} \leq \rho \leq 10^{14} g\cdot cm^{-3}$, a
  few hundred meters.  \newline neutron-rich nuclei in a free gas of
  electrons and neutrons, the neutron pressure increasing progressively
  with density. This is called the free neutron regime.
\item (Outer) core: until $\sim 5\times 10^{14} g\cdot cm^{-3}$, about
  10 km for a pure neutron star (without an inner core) with a mass
  $\sim 1 M_{\odot}$, very variable otherwise. \newline Homogeneous
  liquid core mainly composed of neutrons, with also electrons, protons,
  and muons.
\item Inner core: unknown. Possible appearance of hyperons in the
  hadronic phase, probable transition to deconfined quark matter.
\end{enumerate}

There is also a thin atmosphere (a few centimeters) of atomic gaseous
matter. Only the crust is presently reliably described. The equation of
state of the matter in the crust, being the main constraint in this
work, is studied in more detail in the appendix.  Figure \ref{eoscrust}
shows a plot of the equation of state of the crust.  We take the
standard EOS of the outer crust from the work of Baym, Pethick and
Sutherland (BPS) \cite{BPS} and of the inner crust from the work of Negele and
Vautherin (NV) \cite{Negele73}.

\begin{figure}
\centerline{
\includegraphics{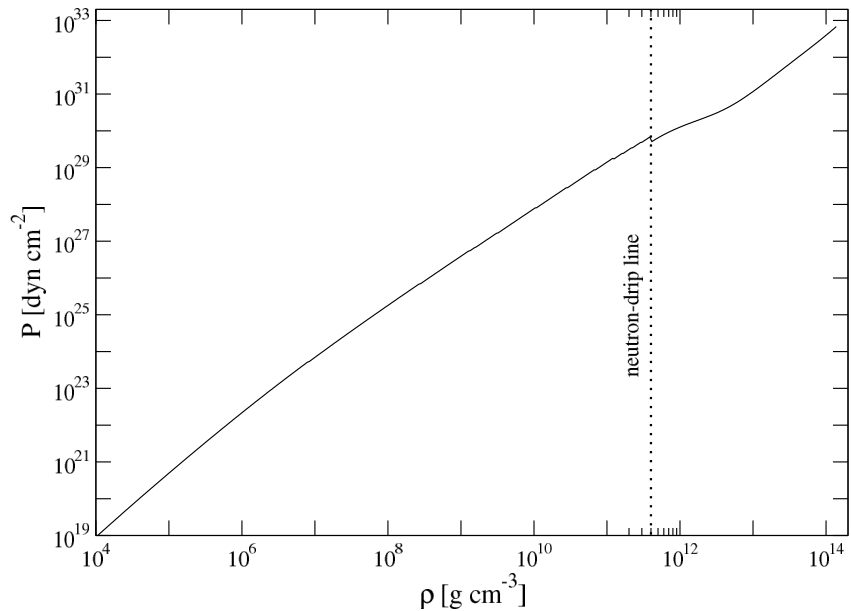}}
\caption{Plot of the `standard' EOS of the crust using the results of
  Baym, Pethick and Sutherland \cite{BPS} and of Negele and Vautherin
  \cite{Negele73}.}
\label{eoscrust}
\end{figure}

We see that the transition to the free neutron regime corresponds to a
kink in the curve: the neutron drip is a first order phase transition of
the matter in the crust. The EOS of the crust ceases to be valid shortly
before half normal nuclear density.

All what happens after the endpoint of the Negele and Vautherin EOS
(determined by the last nucleus they compute in the original article) is
not known for sure. The properties of the hadronic phase, where nuclei
don't exist anymore, and all possible phase transitions, with apparition
of hyperons, deconfined quark matter \dots are strongly model-dependent,
and should thus be considered as unknown (see e.g.\ the most recent
review by Lattimer and Prakash \cite{LP04}).

\section{Parametric approach to the description of the high-density EOS} 
\label{sec:eos}

\subsection{Motivations}

We have seen in the previous section that the EOS of the matter in a
neutron star is very badly known for the hadronic part and the denser
phases. But the mass-radius relation of the neutron star branch, and in
particular the maximum stable mass, is highly dependent on the EOS used
in this domain. Many papers have been issued about neutron stars, each
of them using a different model, and giving different properties for
neutron stars. For a documentation about the many equations of state
available on the market, see e.g. \cite{Weber_book}. To be short, we can
thus say that all models, even the most complicated ones, are
unreliable, and so, why not choose a simple one, but with a certain
number of parameters so as to be able to get a large panel of
mass-radius relations, and to be able to study the influence of these
parameters on the existence of a third family of compact stars, with a
core of ultra-dense matter (which we will assume is quark matter)?

\subsection{Construction of the equation of state}

We constrain all our equations of state by starting them with the BPS
and NV EOS's, then extending continuously from the last point of the NV
EOS, which corresponds to the energy density for the last calculated
nucleus in the original article by Negele and Vautherin, that is $\rho
\simeq 1.35\cdot 10^{14} g\cdot cm^{-3}$. We choose to describe three
other phases after the crust equation of state:

\begin{itemize}
\item a hadronic phase, describing the matter in the outer core, for
  which we adopt a polytropic equation of state.
\item a mixed phase, with both hadronic and deconfined quark matter,
  which we describe with a linear equation of state.
\item a phase of pure quark matter, also described with a linear
  equation of state.
\end{itemize}

We impose also the condition that the matter must be causal, meaning
that the sound velocity must be lower than the light velocity, which we
can write as:
$$\frac{dP}{d\rho}\quad = \quad c_{s}^{2} \leq 1$$
where $c_{s}$ is the sound velocity and with the light speed set to
$c=1$ (natural units).

\subsubsection{Polytropic model}

There are actually several ways to define a polytropic equation of state
(for a most recent discussion see \cite{Herrera04}).
A classical polytropic model for an equation of state is defined as
follows:
$$P = K \cdot \rho^{\Gamma}$$
where $\rho$ is the classical mass
density, and $\Gamma$ is called the adiabatic exponent.  Such an
equation of state is extensively used as a standard to describe white
dwarf material. In our relativistic case, we tried first to replace
directly the classical mass density by the energy density. But if we
proceed so, the equation of state becomes very quickly acausal. We thus
introduced the number density of particles $n$, following Appendix G of
\cite{Shapiro_book}. The mass density is simply $m\cdot n$, with $m$ is
the mean mass of the particles present, and we assume:
$$P = K\cdot n^{\Gamma}$$
The energy density must have the following form:
\begin{equation}
\label{rhoform}
\rho = m\cdot n + \rho'
\end{equation}
$m\cdot n$ being the rest mass energy of the particles, and $\rho'$ being a
term originating from interactions.

Using the first principle of thermodynamics, with the energy per
particle, we get:
$$
d\left(\frac{\rho}{n}\right)\quad = \quad -P d\left(\frac{1}{n}\right) + T ds
$$
where $s$ is the entropy per particle. The matter we consider is
highly degenerate (the Fermi energy is much larger than the typical
temperature), so that we can consider $T = 0$. Integrating
the simple relation between $P$ and $\rho$, we get:
$$\rho = constant\cdot n + \frac{P}{\Gamma - 1}$$
and the constant can
be identified with the mean mass $m$ of the particles by comparison with
equation \ref{rhoform}.  We thus have the equation of state $P =
P(\rho)$ in an analytical parametric form:
\begin{eqnarray}
\label{ppoly}
P &=& K\cdot n^{\Gamma} \\
\label{rhopoly}
\rho &=& m\cdot n + \frac{P}{\Gamma - 1}
\end{eqnarray}
The use of a polytropic model can be interpreted as follows: for any
equation of state, one can define the local adiabatic exponent as:
$$\Gamma = \frac{n}{P}\frac{dP}{dn}$$
$n$ being the baryonic number
density. This definition is of course consistent with the definition of
$\Gamma$ in a polytropic EOS. Adopting a polytropic model for a region
of the EOS is then an approximation consisting in taking $\Gamma$
constant, equal to its average value in the ``true'' EOS.

In the hadronic phase, few electrons remain, and the matter is
essentially composed of neutrons, so that we chose $m = m_{neutron}$.
In the mixed phase, as we don't know exactly which particles are there,
it would be completely arbitrary to give a value for $m$, which
justifies the choice of another model, the simplest being a linear
model.

\subsubsection{Mixed phase}

We choose a linear model for the mixed phase, so as to mimic the general
behaviour found in more sophisticated calculations which take into
account the Gibbs criteria for phase transitions (see \cite{Glen_book}).
The only parameters determining this phase are the energy density at
which it begins, and its slope. We consider a first order phase
transition, so that the slope of this phase must be different from the
slope at the end point of the polytropic phase. We shall moreover assume
that this phase transition leads to a softening of the EOS, that is,
that the slope of the mixed phase is lower than the slope at the end of
the polytropic phase. We thus adopt as a parameter not the slope, but a
factor $f$ between 0 and 1 being the fraction of the slope in the mixed
phase and the slope at the end of the polytropic part. The case $f=0$
applies when the mixed phase is not formed inside the star, e.g. when
the positive surface energy is so large as to prevent the formation of
bubbles.

\subsubsection{Quark phase}

For the innermost part, the quark phase, we based our choice of a linear
model on the success of the phenomenological MIT Bag Model
\cite{Chodos74} which leads to an equation of state of the form:
$$P = \frac{1}{3}\rho - \frac{4}{3}B$$
used as a standard for describing
quark matter cores and strange stars (see e.g.\ 
\cite{Haensel86,Alcock86}) Here $B$ stands for the vacuum energy density
associated with the quark phase, and is a phenomenological constant. In
this work its value will be simply fixed by the constraint of having a
continuous EOS with the mixed phase. A slope of 1/3 is used throughout this
work like in the MIT Bag Model, which corresponds to that one would
find for an ultra-relativistic ideal Fermi gas. Quarks are indeed
expected to behave like such a gas at ultra-high densities, and the
slope is predicted to be lower than 1/3 if there exist interactions
between them, so that this value of 1/3 can be considered as a firm
higher limit for this slope (see \cite{FPS01}).

\begin{figure}
\includegraphics{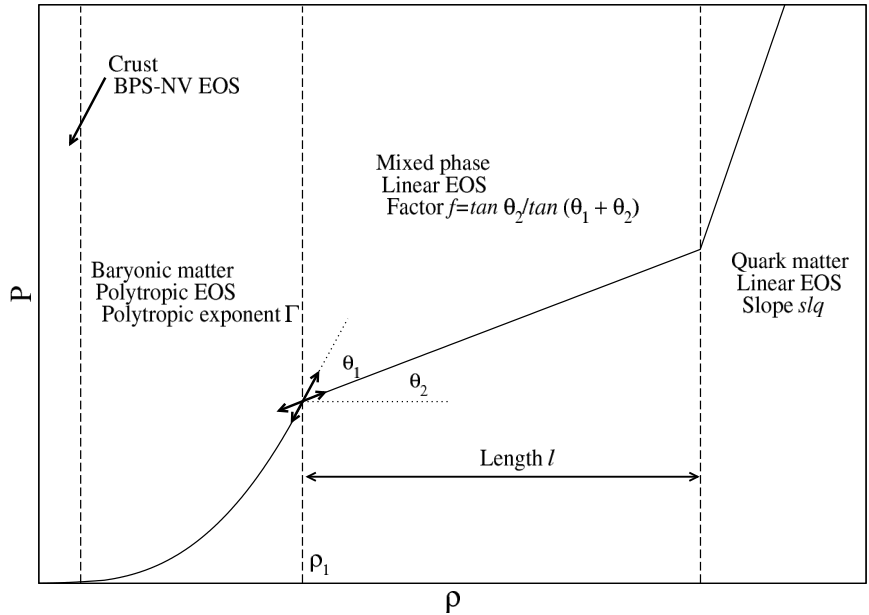}
\caption{Parameters characterizing the EOS studied. Each region is
  described by the nature of the matter present, the model chosen for
  the equation of state and the parameters characterizing this region.}
\label{eosparam}
\end{figure}

Finally, the parameters determining the EOS are:

\begin{enumerate}
\item the polytropic exponent $\Gamma$. Once its value is given, the
  factor K is determined by continuity with the crust EOS.
\item the beginning of the mixed phase, $\rho_{1}$.
\item the factor $f$ for the slope of the mixed phase.
\item the length of the mixed phase $\ell$.
\item the slope of the quark phase, $slq$.
\end{enumerate}

These notations will be used throughout the rest of this report without
recalling their significance.
Figure \ref{eosparam} summarizes the parameterization for the whole EOS
studied in the following.

\section{Results}

\label{sec:results}

\subsection{Technical precisions}

In the following, we shall denote by $\rho_0$ the density of normal
nuclear matter: 
$$\rho_{0} = 0.17 fm^{-3} = 2.8\cdot 10^{14} g\cdot cm^{-3}$$
The BPS
and NV EOS were compiled at the institute by Stefan R\"uster, who
checked the BPS EOS using the latest nuclear models of the binding
energies of neutron-rich nuclei up to the neutron drip-line. As they are
computed numerically and given in a tabular form, the entire equation of
state is given in this form, even though all the parts after the crust
are analytical.  For the points between two discrete values a linear
interpolation is used, which is exact in the mixed phase and the quark
phase, and a very good approximation in the polytropic phase and the
crust EOS, given the short step-size.  The TOV equation was solved with
a fourth order Runge-Kutta algorithm, the program adjusting
automatically the stepsize in $r$ to have at least 2000 points for one
star (the precision on the total radius is then of a few meters over
about 10 km) and less than 10000 for reasonable computational time.  The
stepsize in the central energy density between two different stars is
also dynamically adapted during the calculations to obtain the
mass-radius relation with a variation of about 1\% of the mass between
two points.  The program written to construct the equations of state
checks whether causality is respected, assuring that we don't work with
unphysical EOS's.

\subsection{The less massive stable neutron star}

\begin{figure}
\begin{center}
\includegraphics{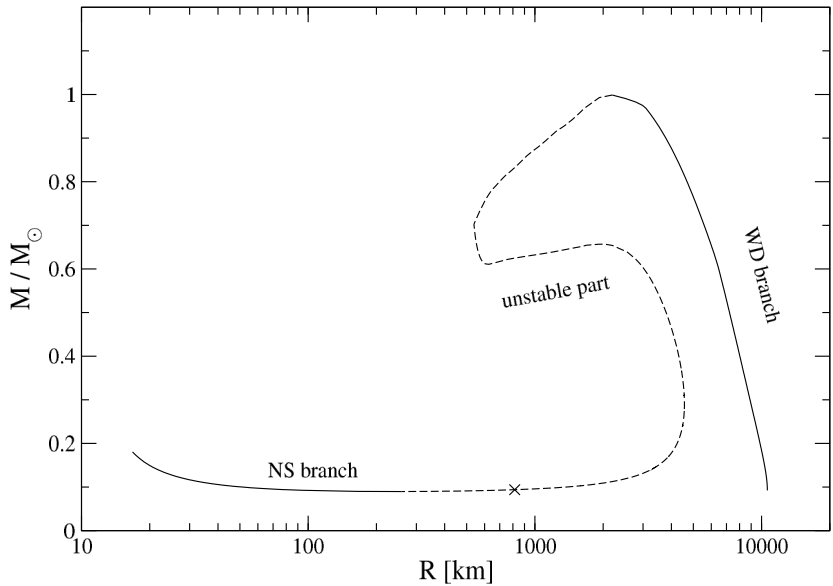}
\end{center}
\caption{Mass-radius relation for central energy densities up to $\rho_{0}$. 
  The almost complete white dwarf branch and the less massive neutron
  stars are visible. The cross indicates the end point of the crust EOS.
  Beyond this point (curve to the left), the stars have a small liquid
  core, described by the polytropic EOS.}
\label{mrcrust}
\end{figure}

Figure \ref{mrcrust} gives the mass-radius relation for stars with a
central energy density between $4\cdot 10^{-10}\rho_{0}$ and $\rho_{0}$.
The equation of state between the end of the crust EOS (just before
$0.5\rho_{0}$) and $\rho_{0}$ is a polytropic one, with $\Gamma = 2.5$.
We verified that the mass-radius relation does not depend on the
polytropic exponent used, when the central energy density stays under
$\rho_{0}$, which can be explained by the fact that most of the star is
described by the crust EOS in this case. On the plot, one can see that
the central energy density increases when we follow the curve from the
right to the left. The regions where $\frac{dM}{d\rho_{c}} < 0$
correspond to unstable stars. It has indeed been shown that this
inequality is equivalent to the instability of the star against radial
oscillations (see \cite{Harrison65,Shapiro_book} and references
therein).

This beautiful plot shows the almost complete white dwarf branch on the
right, with a maximum mass of $1 M_{\odot}$ (the Chandrasekhar mass),
then a long unstable part.  The point on the left where the curve starts
to go up again corresponds to the beginning of the branch of the neutron
stars. The minimum stable mass for a neutron star turns out to be
$M_{min} = 0.09 M_{\odot}$ and the corresponding radius, which is the
largest possible radius for a neutron star, $R_{max} \simeq 260\quad
km$. The corresponding central energy density is approximately
$0.52\rho_{0}$, very close to the energy density at the end of the crust
EOS (0.50$\rho_{0}$), so that the liquid core in this star must be very
small and the neutron star is almost entirely solid. The values found
here are the same that can be found generally in the literature
(\cite{BPS}, \cite{haensel03}).

\subsection{Results for a pure neutron star}

We assume here that no phase transition to quark matter occurs and we
use only the crust equation of state continued by a polytropic equation
of state for the hadronic phase. Under these assumptions, we derive
constraints on the possible values of $\Gamma$.

\subsubsection{Causality: upper physical value of $\Gamma$}

\begin{figure}
\begin{center}
\includegraphics{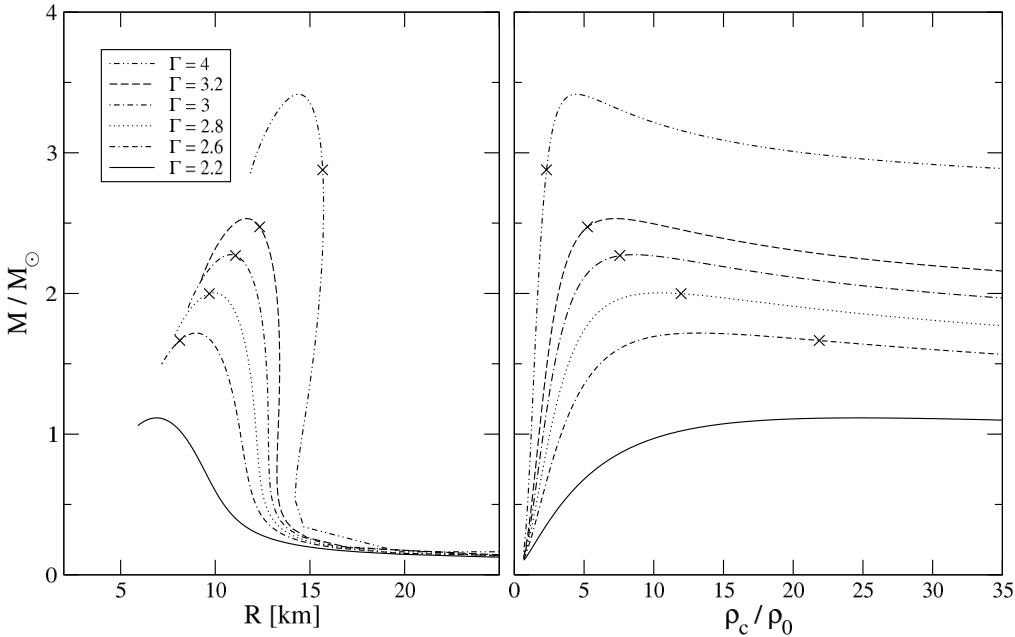}
\caption{Mass-Radius and Mass-Central energy density
  relations for different values of $\Gamma$ for a pure neutron star.
  The crosses on the curves indicate the last causal point. The legend
  is the same for both plots, and indicates the value of $\Gamma$ for
  each curve.}
\label{mrgamma}
\end{center}
\end{figure}

Any polytropic equation becomes acausal in the limit of large densities
if $\Gamma$ is greater than 2. Here we want to describe the
complete mass-radius relation of pure neutron stars with our polytropic
model, that is to say, we want the maximum mass to correspond to a
central energy density where the EOS is still causal. Figure
\ref{mrgamma} shows mass-radius relations for different values of
$\Gamma$ and the corresponding mass-central energy density curves. The
crosses on the curves indicate the point at which causality starts being
violated. We see that the larger $\Gamma$, the smaller the critical
density beyond which causality is violated, until finally it occurs
before the maximum mass is reached. There exists then a limiting value
of $\Gamma$, where the central energy density for the maximum mass
neutron star is precisely equal to the limit of causality.

This figure shows also that the maximum mass increases with $\Gamma$. In
fact, the important point is the \emph{stiffness} of the EOS, which is
related to the slope of the EOS: the greater the average slope, the
higher the maximum mass. The slope increases indeed faster when $\Gamma$
is greater.

\begin{figure}
\begin{center}
\includegraphics{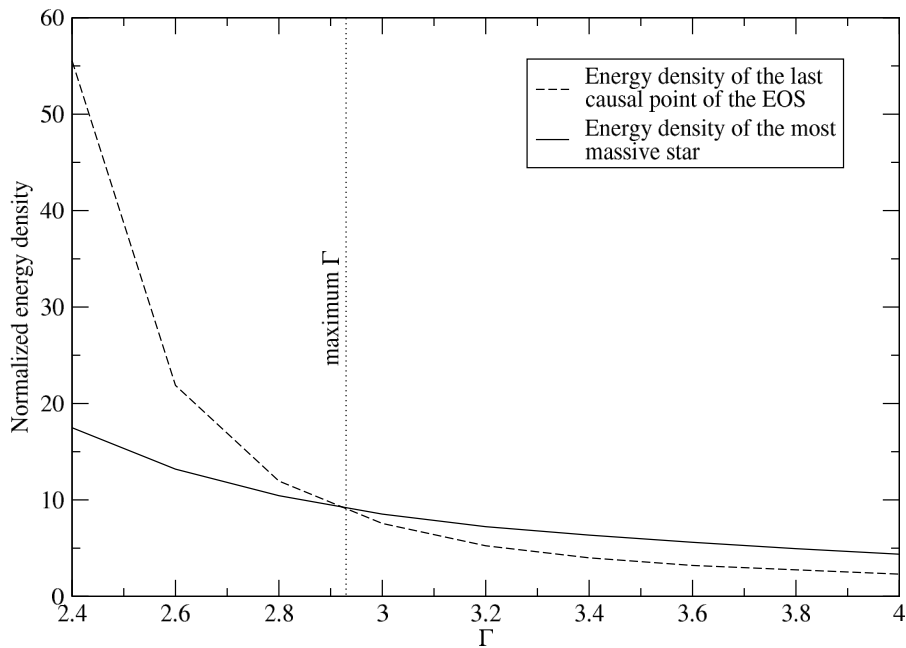}
\caption{This plot shows, for some values of $\Gamma$, the central
  energy density of the most massive pure neutron star and the energy
  density above which the EOS becomes acausal. The intersection of the
  curves gives the maximum value of $\Gamma$ allowed by causality.}
\label{gmemax}
\end{center}
\end{figure}

The figure \ref{gmemax} gives a precise determination of the maximum
$\Gamma$ yielding a causal maximum mass configuration: the energy
density after which the EOS becomes acausal and the energy density
yielding the maximum mass are plotted on the same graph. The abscissa of
their intersection gives the highest physical value of $\Gamma$. We find
$\Gamma_{max} = 2.92$.  The corresponding maximum mass is about $2.2
M_{\odot}$, with a radius of $11 km$. A medium sized star of $1.1
M_{\odot}$ would have a radius of about $13 km$.

\subsubsection{Lower limit for $\Gamma$: maximum observed mass}

\begin{figure}
\begin{center}
\includegraphics{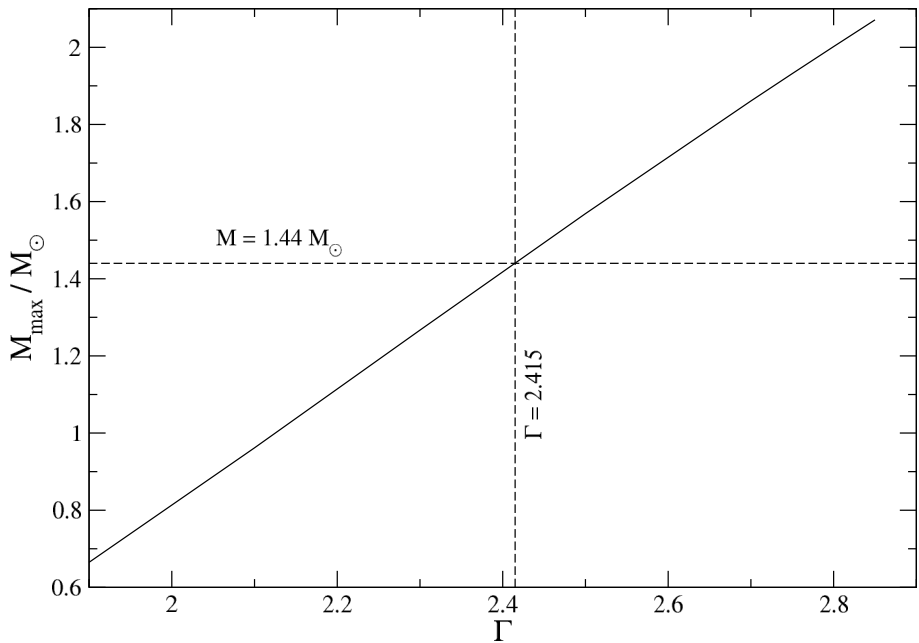}
\caption{This plot shows, for some values of $\Gamma$, the corresponding
  maximum mass. The minimum acceptable value of $\Gamma$ is such that
  the corresponding maximum mass is exactly $1.44\,M_{\odot}$.} 
\label{mmaxgamma}
\end{center}
\end{figure}

We have said that the mass of the Hulse-Taylor pulsar has been precisely
determined to be $1.44 M_{\odot}$, and we want this mass to be included
in our results for the mass-radius relation of pure neutron stars. That
is, we want to obtain a maximum mass greater than $1.44 M_{\odot}$. As
we have seen in the previous section, the maximum mass is an increasing
function of $\Gamma$. The condition
\begin{equation}
\label{hulse}
M_{max} \geq 1.44 M_{\odot}
\end{equation}
gives then a lowest acceptable value of $\Gamma$. Figure \ref{mmaxgamma}
represents the value of the maximum mass as a function of $\Gamma$. We
find that $\Gamma$ must be greater than $\Gamma_{min} = 2.415$ to
satisfy the inequality \ref{hulse}. The radius of the most massive star
is then as small as $8 km$. A medium sized star with $M = .7 M_{\odot}$
would have a radius of about $10 km$.

\subsubsection{Conclusions}

Thus, we find a range of physical values for the average adiabatic
exponent of the equation of state under the assumption that there is no
phase transition to a phase of quark matter.  These constraints don't
remain valid if we consider a phase transition. Indeed, the polytropic
phase ends as soon as the mixed phase begins, and need only remain
causal until this point.

\subsection{Stars with a phase transition}

\subsubsection{New upper limit for $\Gamma$}

This time, we only impose the polytropic part to remain causal until the
beginning of the phase transition. The maximum physical $\Gamma$ will
then depend on the chosen value of $\rho_{1}$. The figure \ref{gmemax}
showed the value of the energy density where the polytropic part of the
EOS becomes acausal, as a function of $\Gamma$. This value is also, for
a given $\Gamma$, the highest possible value for $\rho_{1}$. We assume
that the phase transition doesn't begin before $\rho = 2\rho_{0}$. The
corresponding value of $\Gamma$ on figure \ref{gmemax}, which is the
highest possible value of $\Gamma$, is then $\Gamma=4.2$.

\subsubsection{Effect of $\Gamma$}

\begin{figure}
\begin{center}
\includegraphics{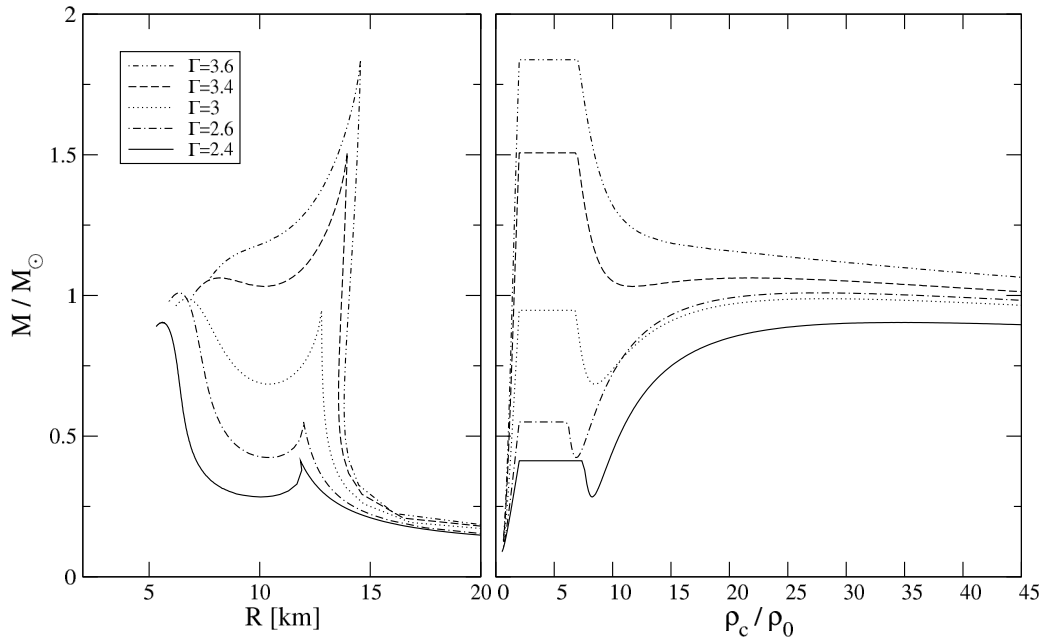}
\caption{Influence of the polytropic coefficient $\Gamma$ of the
  hadronic EOS on the mass-radius and mass-central energy density
  relations. A phase transition to quark matter occurs at $2\rho_0$ with
  a zero slope mixed phase EOS with a length of $3\rho_0$.}
\label{g1_fixed}
\end{center}
\end{figure}

We vary the value of $\Gamma$ while all other parameters have \emph{a
  priori} reasonable fixed values: $\rho_{1} = 2\rho_{0},\quad f =
0,\quad \ell = 3\rho_{0}, \quad slq = 1/3$.  The choice $f = 0$ means
that the slope of the mixed phase is zero. The right-hand side of the
TOV equation is always strictly lower than zero inside the star, so that
the pressure cannot remain constant in a region limited by two different
radii. Hence, there is a jump in energy density in the star, and one
goes directly from the hadronic phase to the quark phase, without a
smooth transition and a mixed phase in between.  Figure \ref{g1_fixed}
shows the resulting mass-radius and mass-central energy density
relations for several values of $\Gamma$.

There are some remarks in order: first, under these conditions, there is
always an instability appearing just after the phase transition, for
small quark matter cores. As soon as the mixed phase starts, the slope
of the mass-radius relation changes its sign ($dM/dR > 0$) signalling
the onset of an instability (see e.g. \cite{Shapiro_book}).  But there
is then a possibility for a third stable family of compact stars to
exist: the slope of the mass-radius relation changes again to the one of
ordinary neutron stars ($dM/dR<0$) starting a new sequence of stable
solutions to the TOV equations. This behaviour is easily understandable,
because of the jump in energy density: a small core of quark matter is
already very massive, and creates a strong gravitational pull, that the
continuously evolving pressure cannot compensate at first; but the
pressure increases rather fast with energy density in the quark phase,
so that when the inner core grows, it can be stabilized against
gravitational collapse at some critical size.

\begin{table}
\begin{center}
\begin{tabular}{c|c|c}
$\Gamma$ & slope at $2\rho_{0}$ & slope at $3\rho_{0}$\\
\hline
2.4 & 8.4594e-002 & 1.3908e-001\\
2.6 & 1.1794e-001 & 2.0376e-001\\
2.8 & 1.6252e-001 & 2.9297e-001\\
3 & 2.2016e-001 & 4.0809e-001\\
3.2 & 2.9513e-001 & 5.5329e-001\\
3.3 & 3.3905e-001 & 6.3960e-001\\
3.4 & 3.9083e-001 & 7.3234e-001\\
3.6 & 5.0583e-001 & 9.3502e-001\\
\end{tabular}
\caption{Slope of the polytropic EOS at 2 and 3$\rho_{0}$}
\label{tabslope}
\end{center}
\end{table}

Second, for sufficiently high values of $\Gamma$, the third family does
not exist anymore. The corresponding critical value for $\Gamma$ is
between 3.4 and 3.6. This can be linked to the difference between the
slope at the end of the polytropic phase and the slope of the quark
phase: table \ref{tabslope} shows the slope of the polytropic phase in
$\rho = 2\rho_{0}$ and $\rho = 3\rho_{0}$ for several values of
$\Gamma$. We see that in the domain between 3.4 and 3.6 where the third
family disappears this slope begins precisely to be slightly higher than
1/3. It seems that the slope of the quark phase must be higher than the
slope at the end of the polytropic phase to ensure a new onset of
stability.

\subsubsection{Effect of the length of the mixed phase}

\begin{figure}
\begin{center}
\includegraphics{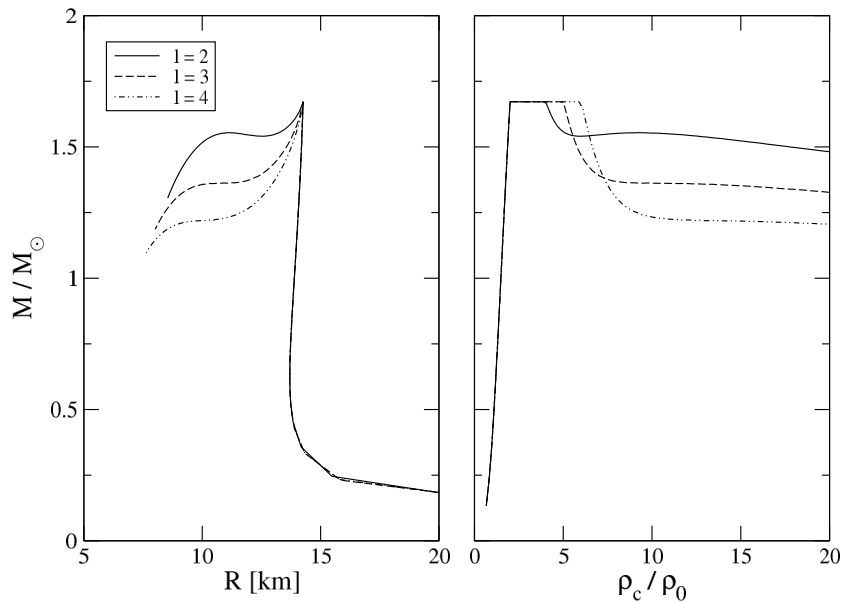}
\caption{Mass-radius and mass-central energy density
  relations for a polytropic value of $\Gamma = 3.5$, with different
  lengths of the mixed phase. The legend indicates the value of the
  length of the mixed phase in $\rho_0$ and is the same for both plots.}
\label{lm}
\end{center}
\end{figure}

Given the previous considerations, we expect the stability of the third
family to be harder to achieve for larger jumps in energy density (i.e. the
length of the mixed phase is larger). Figure \ref{lm} confirms this. The
values of the parameters are the same as before, but we fixed $\Gamma =
3.5$ and we gave the length of the mixed phase three different values,
2$\rho_{0}$, 3$\rho_{0}$, and 4$\rho_{0}$.

The value $\Gamma = 3.5$ seems to be the critical value for the
existence of a third family, when the mixed phase is $3\rho_{0}$ long as
before: there seems to be an inflexion point in both the mass-radius and
the mass-central energy density relations for these values,
corresponding to a metastable point. For a longer mixed phase, there is
no stability at all, but it is interesting to see that if we choose the
region of the mixed phase to be sufficiently small, the case $\Gamma =
3.5$ exhibits a new stable branch, i.e.\ a third family of compact stars.

\subsubsection{Effect of the slope of the mixed phase}

We have studied so far cases with a zero slope for the mixed phase, so
that because of the strictly monotonic increase of the pressure from the
surface towards the center, there is actually no mixed phase inside the
star. There is rather a sudden change of phase, from a pure hadronic
phase to a pure quark phase, accompanied by a jump in energy density,
while the pressure increases continuously. The gravitational pull
increases then a lot compared to the small, continuous increase of the
central pressure. This is the very cause of the instability
systematically observed on the preceding curves, for small quark
cores.

\begin{figure}
\includegraphics{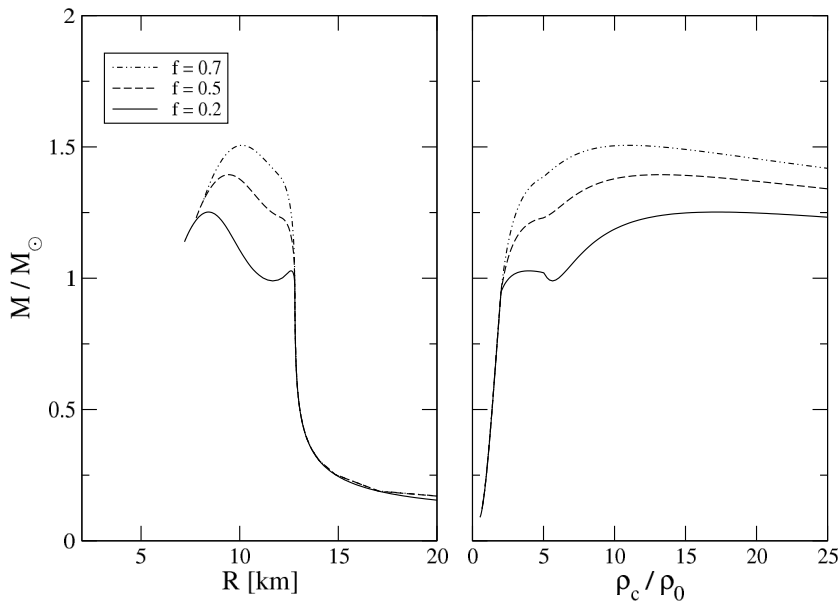}
\caption{Mass-radius and mass-central energy density
  relations for different values of the slope of the mixed phase. The
  legend is the same for both plots. The values of the other parameters
  are fixed and are: $\Gamma = 3$, $\rho_1 = 2\rho_0$, $\ell = 3\rho_0$,
  $slq = 1/3$.}
\label{slm}
\end{figure}

But a general Gibbs condition for the mixed phase gives actually always
a finite slope, so that we consider here non-zero slopes. As said
before, this slope will be given as a factor $f$ between 0 and 1
multiplied by the slope of the polytropic phase at its end.  Figure
\ref{slm} shows the results obtained for $\Gamma = 3$, $\rho_{1} =
2\rho_{0}$, $\ell = 3\rho_{0}$, $slq = 1/3$ and $f = 0.2, 0.5, 0.7$.

These plots show the crucial influence of this slope on the existence of
a third family: for a high enough slope, there is no instability between
the neutron stars and the quark stars, which thus form only one family.
In the case where an instability takes place, we see that the presence
of a non-zero slope for the mixed phase gives a smoother, more realistic
curve than for a zero slope. The mass-central energy density plot shows
that the instability begins for a central energy density in the range of
the mixed phase, so that the most massive neutron stars would have in
this model a hybrid core of hadronic and quark matter.

\subsubsection{Other considerations}

\begin{figure}
\begin{center}
\includegraphics{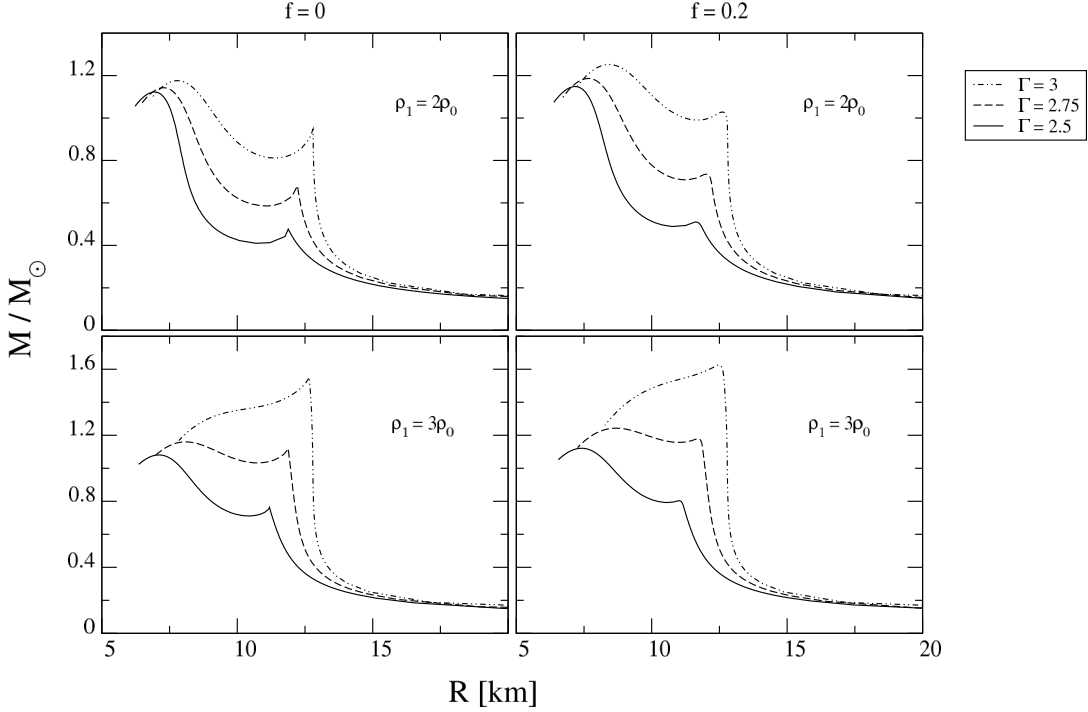}
\caption{Mass-radius relations for various EOS's with a different
  polytropic index. In all plots, $slq=1/3, \ell=3\rho_{0}$. Left plots:
  $f=0$ (no mixed phase), right plots $f=0.2$ (with mixed phase). The
  top and bottom plots show the results for different lengths of the
  mixed phase $\rho_1$.}
\label{other}
\end{center}
\end{figure}

By interpreting the value of $\Gamma$ in terms of microscopic physics,
it is possible to think of additional constraints for its value. The
adiabatic index $\Gamma$ stands in our approach for the power of the
number density $n$ in the interaction term in the energy density, as
seen in equations \ref{ppoly} and \ref{rhopoly}. Indeed, our form of the
EOS used resembles the EOS of the Skyrme or Hartree-Fock model (see
e.g.\ \cite{Shapiro_book}). It is then related to the mean order of the
interactions between the hadrons. The value $\Gamma = 2$ describes a
two-particle interaction, while $\Gamma = 3$ describes a three-body
interaction and so on. It seems then reasonable to say that $\Gamma$
should be between $\Gamma=2\dots3$, the cross section for a more than
$n>3$ n--body interactions being usually considered to be quite small.
Under this new assumption, combined with the requirement $M_{max} \geq
1.44 M_{\odot}$ either in the neutron stars branch or in the quark stars
branch, we can derive some new constraints on the EOS as derived from
the plots in figure \ref{other}.

The left plots are for a constant pressure in the mixed phase (Maxwell
construction or no mixed phase region in the star), while the right
plots take into account a slight increase of the pressure in the mixed
phase, which is actually realized in general in a thermodynamically
consistent treatment (the Gibbs criteria, see \cite{Glen_book}). We
note, that a mixed phase appearing inside the compact star tends to
smoothen out the edges at the maximum mass seen, if there is no mixed
phase present in the compact star. In addition, the masses are slightly
increased due to the additional contribution of the mixed phase to the
pressure.

The upper plots show the mass-radius curves for a mixed phase starting
at $\rho_1 = 2\rho_0$, the lower ones for $\rho_1 = 3\rho_0$.  The
obtained maximum masses are too small, i.e.\ the maximum mass stays
below $M_{max}=1.44 M_\odot$, if the mixed phase starts at $2\rho_{0}$.
If the mixed phase starts at $3\rho_{0}$, a value of $\Gamma$ between
approximately 2.8 and 3.0 is needed to get masses above $1.44
M_{\odot}$, and these masses are obtained only for the neutron stars
branch. A third family appears for smaller values of $\Gamma$. For
$\Gamma=3$, however, the third family does not exist. In order to get a
third family solution and to be compatible with the Hulse-Taylor pulsar
mass, one obviously needs at least some fine-tuning of the parameters of
the EOS used or a different ansatz for the mixed phase and quark phase
EOS.

\subsubsection{Conclusions}

We can now give semi-quantitative ranges of acceptable values for the
different parameters and criteria for the existence of a third family
of compact stars.

\begin{itemize}
  
\item Microscopically, the value for $\Gamma$ should be in the range
  between $\Gamma=2$ and $\Gamma=3$. But the plots \ref{other} suggest
  that it must be near $\Gamma=3$ to get a maximum mass greater than
  $1.44 M_{\odot}$, telling us that three-body or n-body interactions
  are dominant in the hadronic phase. This behaviour is understandable
  given the very high densities involved.
\item The chances for a third family to exist are best when the phase
  transition begins early, because if so, the slope at the end of the
  polytropic EOS is not too big compared to the slope of the quark EOS.
  But if the phase transition occurs too soon for a given $\Gamma$, it
  becomes impossible to get a sufficiently high maximum mass. The
  existence of the third family depends thus significantly on the
  combination of $\Gamma$ and $\rho_{1}$. For $\Gamma$ being between 2.8
  and 3 approximately, $\rho_{1}$ should be more than 2 and less than 3
  according to plots \ref{other}.
\item The existence of the third family depends also crucially on the
  length of the mixed phase: if it is too large, there is no new onset
  of stability after the neutron stars branch, as shown in plot
  \ref{lm}. This plot and figure \ref{g1_fixed} show that the acceptable
  lengths for the mixed phase change significantly with the value of
  $\Gamma$, and can be longer when $\Gamma$ is smaller. More
  calculations would be needed to give a precise range, but values
  between 0 and 4 for $\Gamma$ between 2.8 and 3 seem acceptable, with a
  zero slope for the mixed phase.
\item Figure \ref{slm} suggests that the slope of the mixed phase should
  remain weak to ensure the existence of a third family. Otherwise,
  there is no instability between the neutron star branch and the quark
  star branch. According to plot
  \ref{slm}, the factor $f$ should be between 0 and less than 0.5.
\item All reasonable combinations of parameters yield approximately the
  same ranges of values for the radius: between 8 and 11 km for quark
  stars and from 11 to 260 km with a typical value of about 12 km for
  neutron stars.
\item Finally, we must write a word about the slope of the quark EOS: a
  value of 1/3 is used throughout this work. A smaller value for the
  slope will lower the acceptable values for a third family given above.
  Indeed, the existence of the third family is after all determined by
  the difference between the slope at the end of the polytropic phase
  and the quark phase.
\end{itemize}

We are aware of the unfinished character of this work. The great number
of parameters, and their correlations (changing $\Gamma$ changes also
the difference of slopes between the quark phase and the end point of
the polytropic phase, for instance) makes it tricky to sort out the
results. More refinement in the calculations, and clarification of the
results are needed and are left for future research projects of this
kind.

\begin{acknowledgments}
We thank Matthias Hanauske and Stefan R\"uster for helpful discussion
and for providing the crust EOS and Luis Herrera for instructive
comments.  
\end{acknowledgments}

\appendix*

\section{Equation of state of the crust}

\subsection{Before neutron drip: the BPS equation of state}

\begin{figure}
\begin{center}
\includegraphics[scale = 0.75]{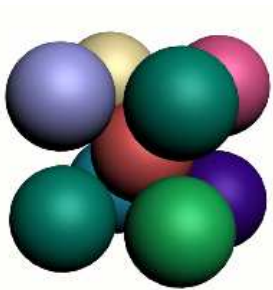}
\caption{Compact view of the geometry of a bcc (body centered cubic)
  lattice} 
\label{bcc}
\end{center}
\end{figure}

The commonly used equation of state for the outer crust was introduced
by Baym, Pethick and Sutherland in 1971 (see \cite{BPS}), and will be
referred to as the BPS EOS. The matter to describe consists of nuclei,
organized in a lattice so as to minimize their Coulomb energy,
surrounded by a free gas of relativistic electrons. Previously to the
work of BPS, it was found that the energetically most favourable lattice
is a bcc (body centered cubic: see figure \ref{bcc}).

At the densities of interest for white dwarfs and neutron star crusts,
the degeneracy pressure of the electrons is huge, and they are captured
by the nuclei, which become more and more neutron-rich as the density
increases. Inversely, the formed neutron-rich nuclei are prevented from
$\beta$-decay because of the very high chemical potential of the
electrons outside. Keeping only the dominant terms, the energy density
can be written as follows:
\begin{equation}
\label{edbps}
\rho_{tot}(A,Z,n_{N},n_{e})\quad = \quad n_{N}(W_{N} + W_{L}) + \rho_{e}(n_{e})
\end{equation}
where $Z$ and $A$ are the number of protons and the number of nucleons
in the nuclei, $n_{N}$ is the number density of the nuclei, $n_{e}$ that
of the electrons, $W_{N}$ is the total energy (including the rest mass
of the nucleons) of an isolated nucleus, but not including any
electronic energy, $W_{L}$ is the lattice Coulomb energy per nucleus,
including the interaction of the nuclei with the homogeneous charge
distribution of the electrons, and $\rho_{e}$ is the electron energy
density. For a bcc lattice, $W_L$ is given by:
$$W_{L} \quad = -1.819620\times\frac{Z^{2}e^{2}}{a}$$
$a$ being the lattice parameter.

The electrons are essentially free, so that their energy density is the
well-known energy of a relativistic ideal Fermi-gas: 
$$\rho_{e} \quad = \quad
\frac{m_{e}^{4}c^{5}}{8\pi^{2}\hbar^{3}}\Big[x(2x^{2}+1)\sqrt{x^{2}+1} -
ln(x+\sqrt{x^{2}+1})\Big]$$
where $$x = \frac{\hbar k_{F}}{m_{e}c}$$
The binding energies defining $W_{N}$ are given in tables, partly from
experimental data, partly from extrapolations from the semi-empirical
mass formula.  The number of parameters can be reduced by introducing
the baryonic number density $n_{b}$, so that 
$$n_{N} = \frac{n_{b}}{A}, \qquad n_{e} = n_{b}\frac{Z}{A}$$
The equation of state is then calculated by finding the nucleus (that is,
the set of $(A,Z)$) that minimizes the total energy density at a given
$n_{b}$. One obtains then both the nucleus present in the crust at a
given number density $n_{b}$ and the value of the energy density at this
same $n_{b}$. The pressure is obtained via the thermodynamic relation
$$P\quad = \quad n_{b}^{2}\frac{\partial(\rho_{tot}/n_{b})}{n_{b}}$$
One thus gets the EOS in the form $P = P(\rho_{tot})$.

\subsection{The neutron drip and beyond}

The above equation of state ceases to be valid at $\sim 4\cdot 10^{11}
g\cdot cm^{-3}$ for the following reason: at this point, the neutrons
present in the nuclei are so weakly bound that they begin to ``drip''
out of the nuclei, so that in the medium between the nuclei a low density
gas of free neutrons begins to appear. The neutron gas density increases little by
little with the total energy density, while the nuclei continue to
become more and more neutron-rich. It is clear that the presence of this
neutron gas will have a strong effect on the remaining nuclei, and the
description by BPS won't be valid anymore. The description of the
nuclear physics beyond neutron drip has been achieved in 1973 by Negele
and Vautherin \cite{Negele73}.

Beyond neutron drip, the semi-empirical mass formula cannot be used
anymore to extrapolate the energy of the nuclei. Indeed, the mass
formula parameters are determined by a very restricted region of nuclear
configurations:

\begin{itemize}
\item  ratio of protons to neutrons higher than 0.6
\item chemical potential of the protons and neutrons in the nucleus
  about -8 MeV
\item zero external pressure on the nuclei
\end{itemize}

In the domain after neutron-drip, named free neutron regime, the
conditions are the following:

\begin{itemize}
\item ratio of protons to neutrons ranges from 0.1 to 0.3
\item the chemical potential of the neutrons in the nucleus, equal at
  equilibrium to this of the free neutrons, ranges from 0 to +20 MeV
\item the pressure of the external free neutron gas becomes significant.
\item instead of being zero, the surface density of the nucleus
  approaches the density of the gas, and the surface is more diffuse.
\end{itemize}

And so a more fundamental theory is necessary to get the energies of the
nuclei, which we shall not describe here. Terms relative to the neutron
gas must also be added in the total energy density. The principle of the
calculations to obtain the EOS is then similar to the BPS EOS by
minimizing the total energy of the system. Negele and Vautherin used an
elaborate Hartree-Fock calculation to describe the whole system (see
\cite{Negele73}). For more recent work, we refer to
\cite{Douchin01,Shen02}. The physics of neutron star crusts is nicely reviewed
by Haensel in \cite{Haensel01}.

\bibliography{all,literat}

\end{document}